\documentstyle[aps,multicol,epsf,psfig]{revtex}
\newcommand{\beq}{\begin{equation}}
\newcommand{\eeq}{\end{equation}}
\newcommand{\beqa}{\begin{eqnarray}}
\newcommand{\eeqa}{\end{eqnarray}}
\newcommand{\ket} [1] {\vert #1 \rangle}
\newcommand{\bra} [1] {\langle #1 \vert}

\newcommand{\mean}[1]{\langle #1 \rangle}

\begin{document}

\title{Optimal cloning of coherent states with
a linear amplifier and beam splitters}
\author{Samuel L.\ Braunstein,$^1$ Nicolas\ J.\ Cerf,$^{2,3}$ 
Sofyan\ Iblisdir,$^2$ Peter\ van Loock,$^1$ and Serge Massar$^4$}
\address{$^1$ Informatics, Bangor University, Bangor LL57 1UT, UK\\
$^2$ Ecole Polytechnique, CP 165, Universit\'e Libre de Bruxelles, 
1050 Brussels, Belgium\\
$^3$ Jet Propulsion Laboratory, California Institute of Technology, 
Pasadena, California 91109\\
$^4$ Service de Physique Th\'eorique, CP 225, Universit\'e Libre de Bruxelles,
1050 Brussels, Belgium\\}

\date{December 2000}
\draft

\maketitle

\begin{abstract}

A transformation achieving the optimal symmetric $N \to M$ cloning of 
coherent states is presented. Its implementation 
only requires a phase-insensitive linear amplifier and a network of
beam splitters.
An experimental demonstration of this continuous-variable cloner 
should therefore be in the scope of current technology.
The link between optimal quantum cloning 
and optimal amplification of quantum states is also pointed out.

\end{abstract}

\pacs{PACS numbers: 03.67.-a, 03.65.Bz, 42.50.Dv, 89.70.+c}

\begin{multicols}{2}

Quantum systems cannot be cloned {\em exactly} \cite{wzd}, but only
approximately.  Finding the optimal {\em approximate} quantum cloning 
transformation has been a fundamental issue in quantum information theory
for the last five years. In quantum cryptography, 
for instance, this problem happens to be strongly related to the
assessment of security \cite{fuch97}. Cloning has been extensively
studied to date for discrete quantum variables, such as quantum bits
\cite{mandel,bh,gisinmassar,brussetal,niu,cerfprl,impl} or $d$-level
systems \cite{werner,bhprl,cerfjmo}, since quantum information theory 
was initially developed for these kinds of systems. 
Recent progress has shown, however, that continuous spectrum systems
might be experimentally simpler to manipulate than their discrete
counterparts in order to process quantum information 
(see, e.g., \cite{qec,telep}).

Stimulated by this progress, we investigate in this Letter
the possibility of achieving the cloning of continuous-variables
quantum information. Commonly, a distinction is made between 
{\em universal} cloning, if the set of input states
contains all possible states for a given Hilbert space dimension,
and {\em state-dependent} cloning, if the input states are restricted to
a certain set which does not contain all possible states.
For any Hilbert space dimension, the optimal universal 
cloner \cite{werner,bhprl,cerfjmo} that clones all possible 
input states equally well can be constructed from a single family of
quantum circuits \cite{Sam}. This universal cloner reduces to a classical
probability distributor in the continuous limit.
Besides the universal cloner, quantum cloning of continuous-variable 
systems has been considered first in a state-dependent context.
In Ref.~\onlinecite{ccl}, the duplication of coherent states
was investigated, and an explicit transformation that is
covariant under displacement and rotation in phase space
was derived. This transformation therefore
clones all coherent states with the same fidelity ($F=2/3$),
although it is not universal, strictly speaking, as its cloning fidelity
is lower for other classes of states such as squeezed states.
The optimality of this continuous-variable cloning transformation
was then proven in Ref.~\onlinecite{gcl}. More generally, 
it was shown that, if one attempts to produce $M$ clones from $N$ original 
replicas of a coherent state $\ket{\alpha}$ ($M \geq N$) with an equal 
fidelity for all $\alpha$'s, the so-called $N$-to-$M$ cloning 
transformation must result in an additional noise 
on both quadratures of each of the $M$ outputs
which has a minimum variance 
\beq\label{optbound}
\overline{\sigma}^2_{N,M}=\left(\frac{2}{N}-\frac{2}{M}\right) 
\Delta x^2_{\rm vac},
\eeq
where the vacuum noise on a quadrature is denoted as 
$\Delta x^2_{\rm vac}=1/2$ ($\hbar=1$). The corresponding maximum
$N\to M$ cloning fidelity is $F_{N,M}=MN/(MN+M-N)$.
However, finding the optimal $N\to M$ cloning transformation and proving
that it actually achieves this maximum fidelity was an open problem.

The present Letter resolves this question. 
We use the Heisenberg picture in order to derive explicitly an 
$N \to M$ symmetric cloning transformation that attains Eq.~(\ref{optbound}).
Remarkably, it appears that implementing this transformation 
only requires a phase-insensitive linear amplifier and 
a set of beam splitters. 
%The paper is organized as follows. First, we 
%define the cloning transformation, and we state its expected properties. 
%Then, we treat the special case of duplication ($N=1,M=2$). Finally, 
%we consider the general case of $N$ identical originals and $M$ clones.
Let $\ket{\Psi}=\ket{\alpha}^{\otimes N} \otimes \ket{0}^{\otimes M-N} 
\otimes \ket{0}_z$ denote the initial joint state of the $N$ input 
modes to be cloned (prepared in the coherent state $\ket{\alpha}$), 
the additional $M-N$ blank modes, and an ancillary mode $z$.
The blank modes and the ancilla are assumed to be initially
in the vacuum state $\ket{0}$. Let $\{x_k,p_k\}$ 
denote the pair of quadrature operators associated with each mode $k$ 
involved in the cloning transformation: $k=0 \ldots N-1$ refers to the 
$N$ original input modes, and $k=N \ldots M-1$ refers to the additional 
blank modes. Cloning can be thought of as some unitary transformation 
$U$ acting on $\ket{\Psi}$, and resulting in a state
$\ket{\Psi''} = U \ket{\Psi}$ such that the $M$ modes are left in
the same (mixed) state which is maximally close to $\ket{\alpha}$.
Alternatively, in the Heisenberg picture, this transformation can be 
described by a canonical transformation acting on the operators $\{x_k,p_k\}$:
\begin{eqnarray}  \label{transformation}
 x''_k  =  U^{\dagger} \; x_k \; U, \quad
 p''_k  =  U^{\dagger} \; p_k \; U, 
\end{eqnarray}
while leaving the state $\ket{\Psi}$ invariant.
We will work in the Heisenberg picture and use the above 
notation throughout this paper, with $x''_k$ denoting the clones 
(i.e., the output modes of the cloning circuit except the ancilla $z$),
because cloning turns out to be much simpler to describe 
from that point of view. We will now impose several requirements
on transformation (\ref{transformation}) that translate the
expected properties for an optimal cloning transformation. 
First, we require that the $M$ output modes have the desired 
mean values: 
\beq
\langle {x_k}''\rangle =\bra{\alpha}{x_0} \ket{\alpha}, \qquad
\langle {p_k}''\rangle =\bra{\alpha}{p_0} \ket{\alpha}, 
\eeq
for $k=0 \ldots M-1$. Roughly speaking, 
this means that the state of the clones is centered on the original
coherent state.
Our second requirement is covariance with respect to rotation in phase 
space. Coherent states have the property that the quadrature variances are 
left invariant by complex rotations in phase space. So, for any input mode 
$k$ of the cloning transformation and for any operator $v_k=c\, x_k+d\, p_k$
(where $c$ and $d$ are complex numbers satisfying $|c|^2+|d|^2=1$), 
the error variance $\sigma^2_{v_k}$ is the same:
\beqa
\sigma^2_{v_k} &=& \mean{(v_k)^2}-\mean{v_k}^2 
=\Delta x^2_{\rm vac}=\frac{1}{2}.
\eeqa
We impose this property to be conserved through the cloning process. 
Taking optimality into account, Eq. (\ref{optbound}),
rotational covariance yields:
\beq   \label{variance-clone}
\sigma^2_{v''_k}=\left(1+\frac{2}{N}-\frac{2}{M}\right)\Delta x^2_{\rm vac},
\eeq
where $v''_k=c\, x''_k+d\, p''_k$.
Our third requirement is, of course, the unitarity of the cloning
transformation (\ref{transformation}). 
In the Heisenberg picture, this is equivalent to demanding that 
the commutation relations are preserved through the evolution:
\beq
[{x_j}'',{x_k}'']=[{p_j}'',{p_k}'']=0, 
\quad
[{x_j}'',{p_k}'']=i \delta_{jk},
\eeq
for $j,k=0\ldots M-1$ and for the ancilla.

Let us first focus on the continuous-variable duplication ($N=1$, $M=2$). 
A simple transformation obeying the three conditions mentioned above 
is given by:
\beqa \label{dupli-xp}
x_0''&=&x_0+\frac{x_1}{\sqrt{2}}+\frac{x_z}{\sqrt{2}}, \qquad
p_0''=p_0+\frac{p_1}{\sqrt{2}}-\frac{p_z}{\sqrt{2}}, \nonumber \\
x_1''&=&x_0-\frac{x_1}{\sqrt{2}}+\frac{x_z}{\sqrt{2}}, \qquad
p_1''=p_0-\frac{p_1}{\sqrt{2}}-\frac{p_z}{\sqrt{2}}, \nonumber \\
x_z'&=&x_0+\sqrt{2}\, x_z, \qquad \qquad ~
p_z'=-p_o+\sqrt{2}\, p_z.
\eeqa
This transformation clearly conserves the commutation rules, 
and yields the expected mean values $(x_0,p_0)$ for the two
clones (modes $0''$ and $1''$). Also, one can check that the 
quadrature variances of both clones are equal to $2\Delta x^2_{\rm vac}$,
in accordance with Eq.~(\ref{variance-clone}). This transformation
actually coincides with the Gaussian cloning machine introduced
in Ref.~\cite{ccl}. Interestingly, we note here that
the state in which the ancilla $z$ is left after cloning
is centered on $(x_0,-p_0)$, that is
the {\em phase-conjugated} state $\ket{\alpha^*}$.
This means that, in analogy with the universal qubit
cloning machine \cite{bh}, the continuous-variable cloner
generates an ``anticlone'' (or time-reversed state) 
together with the two clones.

\begin{figure}[htb]
\begin{center}
\epsfxsize=0.68 \hsize\leavevmode\epsffile{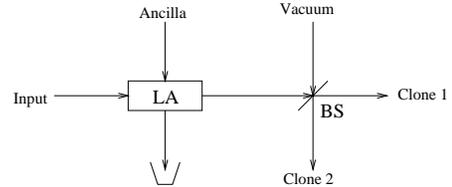}
\end{center}
\narrowtext
\caption{Implementation of a $1 \to 2$ continuous-variable cloning machine. 
LA stands for linear amplifier, and BS represents a phase-free
50:50 beam splitter.} 
\label{fig:clon12}
\end{figure}

Now, let us show how this duplicator can be implemented
in practice. Equation~(\ref{dupli-xp}) can be interpreted as
a sequence of two canonical transformations:
\beqa\label{dupli}
a'_0  &=&  \sqrt{2} a_0 + a_z^{\dagger}, \qquad
a'_z  =  a_0^{\dagger} + \sqrt{2} a_z, \nonumber\\
a''_0 &=&  \frac{1}{\sqrt{2}}(a'_0+a_1), \quad
a''_1 =  \frac{1}{\sqrt{2}}(a'_0-a_1),  
\eeqa
where $a_k=(x_k+ip_k)/\sqrt{2}$ and $a_k^{\dagger}=(x_k-ip_k)/\sqrt{2}$ 
denote the annihilation and creation operators for mode $k$. 
As shown in Fig.~\ref{fig:clon12}, the interpretation of this transformation 
becomes then straightforward: the first step (which transforms $a_0$ and $a_z$
into $a_0'$ and $a_z'$) is a phase-insensitive amplifier whose (power) 
gain $G$ is equal to 2, while the second step (which transforms $a_0'$
and $a_1$ into $a_0''$ and $a_1''$) 
is a phase-free 50:50 beam splitter \cite{capri}. 
Clearly, rotational covariance is guaranteed here by the use of
a {\em phase-insensitive} amplifier. 
As discussed in Ref.~\onlinecite{cave}, the ancilla $z$ involved in linear 
amplification can always be chosen such that $\mean{a_z}=0$, so that
we have $\mean{a''_0}=\mean{a''_1}=\mean{a_o}$ as required. Finally, 
the optimality of our cloner can be confirmed from known results
on linear amplifiers. For an amplifier of 
(power) gain $G$, the quadrature variances of $a_z$ 
are bounded by~\onlinecite{cave}:
\beq\label{amplibound}
\sigma^2_{a_z} \geq (G-1)/2.
\eeq
Hence, the optimal amplifier of gain $G=2$ yields 
$\sigma^2_{a_z} =1/2$, so that our cloning transformation is optimal.

\begin{figure}[htb]
\begin{center}
\epsfxsize=1 \hsize\leavevmode\epsffile{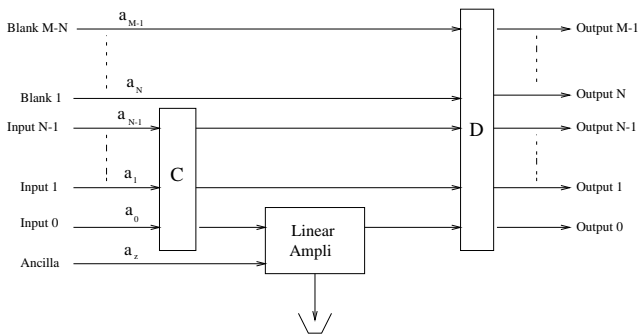}
\end{center}
\narrowtext
\caption{Implementation of an $N \to M$ continuous-variable
cloning machine. C stands for the amplitude concentration operation, while
D refers to the amplitude distribution. Both can be achieved by using
a DFT, or, alternatively, an inverse ``$N$-splitter'' and an 
``$M$-splitter'' (in which case we shift the indices by one in the text,
i.e., $k=1 \ldots M$).} 
\label{fig:clonnm}
\end{figure}

Let us now consider the $N \to M$ cloning transformation.
In order to achieve cloning, energy has to be 
brought to the $M-N$ blank modes in order to drive them from the vacuum 
state into a state which has the desired mean value. We will again achieve 
this operation with the help of a linear amplifier. 
From Eq. (\ref{amplibound}), 
we see that the cloning induced noise essentially originates from the 
amplification process, and grows with the amplifier gain. So, we should
preferably amplify as little as possible. Loosely speaking, the cloning 
procedure should then be as follows: (i) symmetrically amplifying
the $N$ input modes by {\em concentrating} them into one single mode, 
which is then amplified; (ii) symmetrically {\em distributing} the output of 
this amplifier among the $M$ output modes. As we will see,
a convenient way to achieve these concentration and distribution processes
is provided by the Discrete Fourier Transform (DFT). Cloning is then
achieved by the following three-step procedure (see Fig.~\ref{fig:clonnm}).
First step: a DFT (acting on $N$ modes),
\beq
a'_k=\frac{1}{\sqrt{N}} \sum_{l=0}^{N-1} \exp(ikl 2\pi/N) \; a_l,
\eeq
with $k=0\ldots N-1$. 
This operation concentrates the energy of the $N$ input modes into one single 
mode (renamed $a_0$) and leaves the remaining $N-1$ modes 
($a'_1 \ldots a'_{N-1}$) in the vacuum state. 
Second step: the mode $a_0$ is amplified with a linear amplifier 
of gain $G=M/N$. This results in
\beqa
a'_0&=&\sqrt{\frac{M}{N}} \; a_0 + \sqrt{\frac{M}{N}-1} \; a_z^{\dagger}, 
\nonumber \\
a'_z&=&\sqrt{\frac{M}{N}-1} \; a_{0}^{\dagger}+\sqrt{\frac{M}{N}} \; a_z.
\eeqa
Third step: amplitude distribution by 
performing a DFT (acting on $M$ modes) between the mode $a'_0$ 
and $M-1$ modes in the vacuum state:
\beq
a''_k=\frac{1}{\sqrt{M}} \sum_{l=0}^{M-1} \exp(ikl 2\pi/M) \; a'_l,
\eeq
with $k=0\ldots M-1$, and $a'_i=a_i$ for $i=N \ldots M-1$. 
The DFT now distributes the energy 
contained in the output of the amplifier among the $M$ output clones. 

It is readily checked that this procedure meets our three requirements, 
and is optimal provided that the amplifier is optimal, that is    
$\sigma^2_{a_z}=[(M/N)-1]/2$. The quadrature variances of the $M$
output modes coincide with Eq.~(\ref{variance-clone}).
As in the case of duplication, the quality of cloning decreases 
as $\sigma^2_{a_z}$ increases, 
that is cloning and amplifying coherent states are two
equivalent problems. For $1 \to 2$ cloning, we have seen that
the final amplitude distribution among the output clones is achieved with a
single beam splitter. In fact, any unitary matrix such as the DFT
used here can be realized with a sequence of beam splitters 
(and phase shifters)\cite{reck}. This means that the $N\to M$ cloning
transformation can be implemented using only passive elements
except for a single linear amplifier.

We will now explicitly give the {\it simplest} beam splitter combination
that enables the above transformation.
For convenience, let us now use the indices $k=1 \ldots N$ for the 
$N$ original input modes $a_k$, and $k=N+1 \ldots M$ for the additional 
blank modes $a_k$.
With an ideal (phase-free) beam
splitter operation acting on two modes $c_k$ and $c_l$,
\beq\label{generalBS}
\left( \begin{array}{c} c_k' \\ c_l' \end{array} \right)=
\left( \begin{array}{cc} \sin\theta &
\cos\theta \\ \cos\theta &
-\sin\theta \end{array} \right) 
\left( \begin{array}{c} c_k \\ c_l \end{array} \right) \;,
\eeq
we define a matrix $B_{kl}(\theta)$
which is an $M$-dimensional identity matrix with the entries
$I_{kk}$, $I_{kl}$, $I_{lk}$, and $I_{ll}$ replaced by the corresponding
entries of the above beam splitter matrix. 
Now we can define a sequence of beam splitters acting on $M$
modes (``$M$-splitter'' \cite{telep}) as 
\beqa\label{msplit}
{\mathcal U}(M)&\equiv&
B_{M-1\,M}\left(\sin^{-1}\frac{1}{\sqrt{2}}\right)B_{M-2\,M-1}
\left(\sin^{-1}\frac{1}{\sqrt{3}}\right) \nonumber\\ &&\times\cdots\times
B_{12}\left(\sin^{-1}\frac{1}{\sqrt{M}}\right) \;. 
\eeqa
The individual beam splitters in Eq.~(\ref{msplit}) depend only on 
their reflectivity/transmittance parameter $\theta$.
In order to concentrate the $N$ identical inputs, we send them now
through an inverse $N$-splitter, 
\beq\label{inverseN}
\left(\begin{array}{cccc} a'_1 & a'_2
& \cdots & a'_N \end{array}\right)^T =
{\mathcal U}^{\dagger}(N)
 \left(\begin{array}{cccc} a_1 & a_2
& \cdots & a_N \end{array}\right)^T . 
\eeq
Again, we end up with one mode (renamed $a_1$) having non-zero mean 
value and $N-1$ modes ($a'_2 \ldots a'_N$) in the vacuum state.
After amplifying mode $a_1$,
$a'_1=\sqrt{M/N}\,a_1 + \sqrt{M/N-1}\,a_z^{\dagger}$, etc.,
a final $M$-splitter operation yields the output clones: 
\beq\label{M}
\left(\begin{array}{cccc} a''_1 & a''_2
& \cdots & a''_M \end{array}\right)^T =
{\mathcal U}(M)
 \left(\begin{array}{cccc} a'_1 & a'_2
& \cdots & a'_M \end{array}\right)^T , 
\eeq
with $a'_i=a_i$ for $i=N+1 \ldots M$.

Since the amplification produces extra noise, our
cloning circuits used as little amplification as possible.
However, rather surprisingly, by first amplifying each
input copy $k=1 \ldots N$ individually,   
\beqa
a'_k&=&\sqrt{\frac{M}{N}} \; a_k + \sqrt{\frac{M}{N}-1} \; 
a_{z,k}^{\dagger}, \nonumber \\ 
a'_{z,k}&=&\sqrt{\frac{M}{N}-1} \; a_{k}^{\dagger}+
\sqrt{\frac{M}{N}} \; a_{z,k}, 
\eeqa
a circuit can also be constructed that
yields optimum fidelities. In the next step, the amplified modes
are {\it each} sent together with
$M-1$ vacuum modes $b_{k,1}$, $b_{k,2}$,...,$b_{k,M-1}$ through
an $M$-splitter 
\beqa\label{M2}
&&\left(\begin{array}{cccc} a'_{k,1} & a'_{k,2}
& \cdots & a'_{k,M} \end{array}\right)^T = \nonumber \\
&&\qquad {\mathcal U}(M)
 \left(\begin{array}{cccc} a'_k & b_{k,1}
& \cdots & b_{k,M-1} \end{array}\right)^T . 
\eeqa
The $N M$ output modes after this operation can be written as
\begin{equation}\label{NtoM2}
a'_{k,l}=\frac{1}{\sqrt{N}}a_k+
\sqrt{\frac{M-N}{MN}}a_{z,k}^{\dagger}+d_{k,l} \;,
\end{equation}
where $l=1 \ldots M$. The noise in each $M$-splitter output
coming from the $M-1$ vacuum inputs is represented by 
mode $d_{k,l}$ having zero mean value and quadrature variances 
of $(M-1)/2M$.
The final step now consists of $M$ inverse $N$-splitters acting on
all modes with the same index $l$, i.e., the $N$ modes $a'_{k,1}$,
and the $N$ modes $a'_{k,2}$, etc.  
The output modes at each $N$-splitter,
\beqa\label{NtoM3}
&&\left(\begin{array}{cccc} a''_l & e_{1,l}
& \cdots & e_{N-1,l} \end{array}\right)^T = \nonumber \\
&&\qquad
{\mathcal U}^{\dagger}(N) \left(\begin{array}{ccccc} a'_{1,l} &
a'_{2,l} & \cdots & a'_{N,l} \end{array}\right)^T , 
\eeqa
contain only noise except for one mode,
\begin{equation}\label{NtoM4}
a''_l=\sum_{k=1}^N \left(\frac{1}{N} a_k+
\sqrt{\frac{M-N}{MN^2}} a_{z,k}^{\dagger}+
\frac{1}{\sqrt{N}} d_{k,l}\right) \;.
\end{equation}
Again, all $M$ clones are optimal, although additional
noise has been introduced at the intermediate steps which
results in $M (N-1)$ ``waste'' output modes.
However, this particular circuit points out that
$N \to M$ cloning of coherent states is effectively a ``classical
plumbing'' procedure distributing classical amplitudes.

Finally, we note that for squeezed-state inputs rather than
coherent states, the transformations and circuits presented
require all auxiliary vacuum modes (the blank modes and the ancillary 
mode $z$) be correspondingly squeezed in order to maintain optimum 
cloning fidelities. This means, in particular, that the amplifier 
mode $z$ needs to be controlled
which requires a device different from a simple phase-insensitive
amplifier, namely a two-mode parametric amplifier.
One can say that the cloning machine capable of optimum cloning 
of all squeezed states with {\it fixed} and {\it known} squeezing 
then operates in a non-universal fashion with respect to all
possible squeezed states at the input \cite{ccl,gcl}.

In summary, an optimal $N$-to-$M$ continuous-variable cloning transformation
for coherent states has been derived, which attains the maximum cloning
fidelity $F_{N,M}=MN/(MN+M-N)$. A possible experimental implementation 
of this cloner has been proposed. We trust that 
this implementation should be achievable with current
technology since it only requires a single linear amplifier 
and $N+M-2$ beam splitters. In Ref.~\onlinecite{dariano}, 
an alternative one-to-two cloning scheme has been proposed 
based on three non-degenerate
optical parametric amplifiers, and its experimental realization is
currently underway at Roma University.
Finally, we pointed out the link between the quality of the best cloner 
and the minimum noise induced by the amplification of a quantum state,
emphasizing that spontaneous emission is here again the mechanism 
that prevents the perfect cloning of quantum states 
of light \cite{mandel,impl}.

S.I. thanks Gilles Van Assche for helpful comments, and 
acknowledges support from the Fondation Universitaire Van Buuren at the 
Universit\'e Libre de Bruxelles. S.M. is a research associate 
of the Belgian National Fund for Scientific Research. 
N.C. and S.M. acknowledge funding by the European Union 
under the project EQUIP (IST-FET programme).
S.L.B. and P.v.L. are funded in part under project QUICOV under the
IST-FET-QJPC programme. P.v.L. acknowledges support by a DAAD
Doktorandenstipendium (HSP III).

\end{multicols}

\end{document}